# The Barycentric Fixed Mass Method for Multifractal Analysis

Y. Kamer[1, 3], G. Ouillon[2], D. Sornette[3]

[1]Swiss Seismological Service, ETH Zürich, Switzerland
[2]Lithophyse, 4 rue de l'Ancien Sénat, 06300 Nice, France
[3]Chair of Entrepreneurial Risks, Department of Management, Technology and Economics, ETH Zürich, Switzerland

**Abstract**:

We present a novel method to estimate the multifractal spectrum of point distributions. The method incorporates two motivated criteria (*barycentric pivot point selection* and *non-overlapping coverage*) in order to reduce edge effects, improve precision and reduce computation time. Implementation of the method on synthetic benchmarks demonstrates the superior performance of the proposed method compared with existing alternatives routinely used in the literature. Finally, we use the method to estimate the multifractal properties of the widely studied growth process of *Diffusion Limited Aggregation* and compare our results with recent and earlier studies. Our tests support the conclusion of a genuine but weak multifractality of the central core of DLA clusters, with $D_q$ decreasing from 1.75±0.01 for q=-10 to 1.65±0.01 for q=+10.

**1-Introduction**

Since their popularization by Mandlebrot [1], fractals and fractal geometry have been empirically observed and extensively studied in a wealth of natural and experimental physical phenomena. A common way to quantify the fractal or multifractal properties of a given set of data points is to calculate its generalized (Renyi) dimensions [2], given as:

$$D_q = \lim_{\varepsilon \to \infty} \frac{\frac{1}{1-q}\log(\sum_i p_i^q)}{\log(\frac{1}{\varepsilon})} \quad (1)$$

where ε is the scale of observation, $p_i(\varepsilon)$ is the fraction of data points (e.g, estimated measure) within box $i$ of size ε, $q$ is a real-valued moment order and the sum is performed over all boxes covering the data set under investigation. The most popular generalized dimensions are: $D_0$ the box counting dimension, $D_1$ the information dimension, and $D_2$ the correlation dimension. Varying the $q$ parameter, $D_q$ characterizes the scaling of the underlying measure within the distribution. Thus, $D_{-\infty}$ and $D_{\infty}$ respectively correspond to the local scaling of the lowest and highest densities, i.e. to the weakest and strongest singularities. For monofractal sets, $D_q$ is a constant independent of $q$. For multifractal distributions, $D_q$ decreases monotonically with $q$, and the resulting functional dependence of $D_q$ as a function of $q$ fully characterizes the underlying scaling properties. However, in practical implementations, strong departures from the theoretical values may occur due to edge effects related to the shape of the sampled zone, or to the finite number of data points (see for instance [3]).

In many studies, researchers have tried to account for the edge and finite size effects using preprocessing, filtering and exclusion of data [4–6], which often suffer from some arbitrariness. The results' sensitivity to these subjective choices is often ignored or deemed incomputable, since these choices often alter not only the applied method but also the data used. Although methods for assessing and correcting such bias have been formerly introduced (see [3,7,8] for instance), some recent fractal analysis studies continue to use methodologies which exhibit errors up to 0.15 in $D_0$ for uniform 2D distributions [9]. Such large error margins and the need for additional corrections to obtain unbiased measures have hindered the interpretation and comparison of the results between different multifractal analyses. In this study, we address the issue of edge effects by introducing a novel method that accounts for such errors/biases intrinsically during the analysis. This is done with the help of two data-driven, non-arbitrary criteria: barycentric pivot selection and non-overlapping coverage.

We test the performance of the method on synthetically generated monofractal and multifractal distributions and compare the obtained empirical results with the analytically predicted ones. Encouraged by the results, we then proceed with the analysis of large clusters resulting from the growth process of diffusion limited aggregation. We provide new results that further inform the debate about the possible multifractal nature of such a generic growth process.

## 2-The Barycentric Fixed Mass Method

### 2.1 *Review of multifractal analysis methods*

In order to put our proposed method in perspective, we shall first give a brief overview of the commonly used multifractal analysis methods. Generally, they are classified as either fixed-size or fixed-mass methods. Fixed-size methods (FSMs) [10,11] estimate $D_q$ via the scaling of the total mass $M$ within a constant $r$-sized ball, as $r$ is increased:

$$\log\left\langle M(<r)^{q-1}\right\rangle \approx (q-1)D_q \log(r) \qquad (2)$$

The box counting method, which consists in covering the distribution with boxes and increasing their sizes, is a classical example of FSMs. Due to its significant bias for small samples, the box counting method is regarded as impractical [12]. Inspired from the correlation dimension algorithm [8], the sand box method [11] performs better than box-counting by centering circles at arbitrary points on the fractal and averaging the mass accumulation as the radii is increased. However, this method is not reliable for $D_q$ values when $q<1$, which quantify the scaling properties of the weakest singularities, i.e. the low density parts of the multifractal.

On the other hand, fixed-mass methods (FMMs) estimate $D_q$ via the scaling of the smallest radius $r$ to include a fixed mass $m$, as $m$ is increase. Several studies report FMMs to be superior to FSMs, especially for negative $q$ values [13–15]:

$$\log \left\langle R(<m)^{-(q-1)D_q} \right\rangle \approx -(q-1)\log(m)$$
$$\tau(q) = (q-1)D_q \qquad (3)$$
$$\log \left\langle R(<m)^{-\tau(q)} \right\rangle^{1/\tau(q)} \approx \log(m)^{-1/D_q}$$

For a detailed review of both FSMs and FMMs, the reader is referred to Theiler [16]. Other methods such as wavelet analysis have also been introduced [17,18]; however they are also prone to biases due to finite size effects, and their efficient implementation generally necessitates to discretize the underlying distribution.

The barycentric fixed-mass method (BFM) introduced in the present study uses Equation (3) to estimate $D_q$. The method uses two criteria in order to reduce the finite size and boundary effects, which we now describe.

*2.2 Barycentric Pivot Selection*

In both FSM and FMM, the data points serving as centers for the fixed radius or mass circles are chosen randomly within the sample. The $Dq$'s measured using a small selection of such random centers is considered to be good approximation if those centers (hereafter pivot points) are chosen according to a uniform distribution on the fractal [11]. This assumption reduces the computation load and allows a quick analysis of large datasets. However the measured multifractal spectrum will depend on the location of the randomly selected pivot points, as the finite size and irregular boundaries effects will vary: pivots in the inner core of the fractal will accumulate more mass compared to pivots on the outer edges. Repeating the analysis with a different set of pivot points will result in variation of the estimated $D_q$, which controls the precision of the analysis. Using all data points as pivots would give a single $D_q$ estimate, increasing the precision, but this would not account for the edge effects and would require more computational resources.

The barycentric pivot selection criterion tackles these two issues and is illustrated in Figure 1. We consider a given data point (plotted in yellow color in Figure 1) as a potential pivot. As the mass $m$ has been previously defined, we consider its $m$ closest neighbors and compute the barycenter of those $m$ datapoints; we also compute $r$, which is the distance from the pivot to the farthest of those neighbors. If the barycenter stands closer to the pivot than to any neighbor, then the corresponding couple *(m, r)* contributes to the averaging term in Equation (3). This is the case in Figure 1 for the circles labeled as *A* and *B*, as well as for the corresponding barycenters of the enclosed data points labeled the same way. Those circles correspond respectively to masses *m=*5 and 10 and radii $r_A$ and $r_B$.

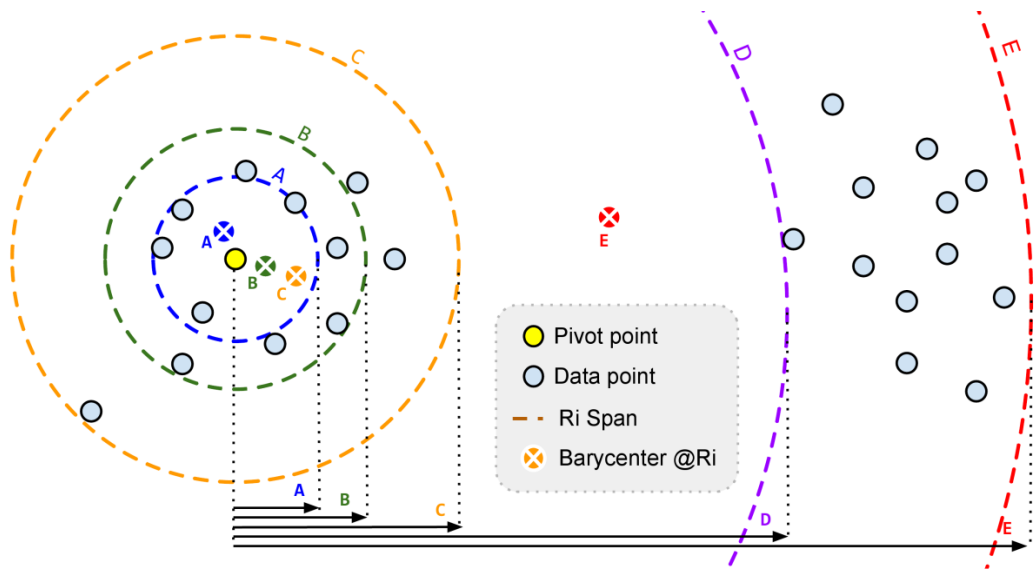

**Figure 1.** Illustration of the barycentric pivot point selection criterion

In the example shown in Figure 1, this criterion then ceases to be valid when one extends the mass $m$ by 3 units or more, as the circle radius now becomes equal to $r_C$ (corresponding to the circle labeled $C$ on the figure). The corresponding barycenter is found to be offset to the right so that the previous pivot point ceases to be active, as it is no longer the closest point to the barycenter. Another example is shown with a circle and its center both labeled as $E$, corresponding to a mass $m=25$. In the usual methods, be it FSM or FMM, the circle around the yellow point is allowed to extend till it encloses all its neighbors. In contrast, using the barycentric criterion, the pivot point will be active only up to radius $r_B$ in the example shown in Figure 1. Applying this criterion to each data point, we determine the set of radii for which it can be used as a pivot.

*2.3 Non-Overlapping Coverage*

Implementing the classical methods, all $N$ pivot points spread over the whole self-similar set, so that each data point contributes $N$ times to the averaging term in Equation (3). The barycentric pivot selection criterion we introduced above results in pivot points being preferably chosen within dense areas where the mass concentration is higher than in their neighborhood. Points located within these areas will be more likely to satisfy the barycentric condition over large radii, resulting in high density areas having a higher contribution to the averaging term. To account for the bias that could result from this selection, we introduce an additional non-overlapping coverage (NOC) criterion for each pivot point. For each fixed mass value, we require that the ensemble of selected pivot points and their respective fixed mass spheres define a non-overlapping configuration. In the absence of the implementation of the NOC condition, the data points located in dense regions would be multiply counted by many spheres, leading to an oversampling of the strongest singularities. By introducing the NOC criterion, we are effectively equalizing the probability of low-density areas to be correctly sampled. The NOC criterion is akin to the construction of the packing dimension, which is obtained by "packing" equal sized spheres inside a given subset [19]. However, because our method is a fixed-mass method, the spheres will have different sizes, contrary to the usual definition of the packing dimension. Allowing for only a limited amount of overlap in the location of the spheres will lead to a tight covering of the dataset, a configuration similar to the definition of the Hausdorff dimension [20]. The minimum amount of overlap can be approximated by considering a set of three circles covering an area as given in Figure 2. Once the radius of the circle is fixed, we compute the minimum distance between the centers of the circles as the one for which all three circles intersect at the same location (ensuring that the whole space within the dashed triangle is completely covered by the three circles). By simple geometrical reasoning applied to the case of three circles of equal radii, we estimate the radius overlap as $2R/2r = 1-\sqrt{3}/2 = 13.4\%$, where r is the radius of the circles and 2R the distance between two distinct circle centers.

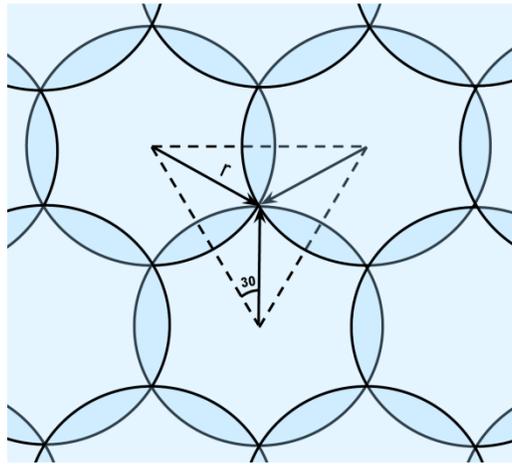

**Figure 2.** Minimum overlap of circles covering an area

The NOC criterion is implemented by placing a first random candidate pivot point with a circle with radius $R_0$ (satisfying the barycentric pivot selection) and then discarding all pivot points (with radii $R_i$) within a distance of $0.86(R_0+R_i)$. By downscaling the radii to 86.6%, we tend to induce an overlap of 13.4% necessary for full coverage. The next pivot point is again chosen randomly from within the remaining set of possible pivot points. The random selection and consequent discarding is then repeated over the remaining set of points until all candidate pivot points are placed (or discarded). For a synthetic dataset we use the multifractal Sierpinski measure which is obtained by recursive replication of the density matrix [1 0; 1 2]. Figure 3 illustrates the first recursion of the generation procedure. At each recursion, the output grid replicates itself multiplicatively over each element of the density matrix. This results in tripling (due to the 3 non zero elements) of the area and quadrupling (1+0+1+2=4) of the mass with each recursion. The reader is referred to [21] for details. The distributions in Figure 4 are obtained by 6 recursive replications, resulting in a total mass of $(1+0+1+2)^6=4096$ points and a maximum mass concentration of $2^6=64$ points. The same figure displays two coverages of a multifractal Sierpinski triangle with fixed-mass circles with masses of respectively 84 and 136. The candidate pivot points satisfying the barycentric pivot selection are plotted as gray dots and the selected circle centers are plotted as black dots.

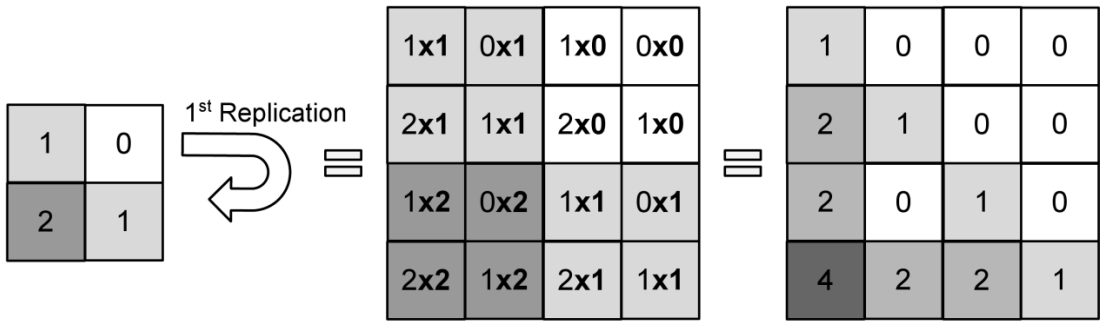

**Figure 3.** Generation of the synthetic multifractal measure shown in Figure 4 from the density matrix given by the 2x2 table on the left

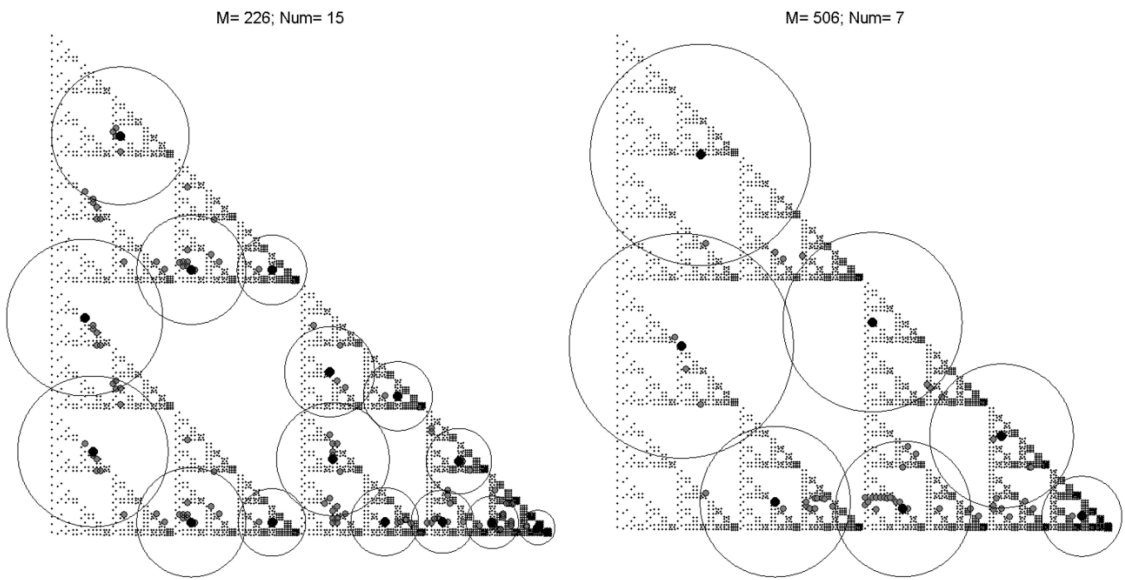

**Figure 4.** Coverage of a synthetic multifractal Sierpinski triangle with fixed-mass circles

### 3-Multifractal spectra of synthetic datasets

#### *3.1 First example with the synthetic multifractal Sierpinski triangle of Figure 4*

To estimate $D_q$, we increment $\tau$ and calculate the expression $\log\langle R(<m)^{-\tau}\rangle^{1/\tau}$ for sets of fixed-mass circles covering the point distribution. The mass range $m$ is sampled at logarithmically spaced steps, giving $m_i = m \times 10^{i\alpha}$, with $\alpha=0.05$, where the role of the smallest value m is discussed in section 3.2. The curves of averaged radii versus fixed-mass are given on a log-log plot in Figure 5a. Calculating the slope for each $\tau$ exponent (represented in shades of gray), we estimate both $D_q$ and $q$. The analytical and estimated curves of $D_q$ as a function of $q$ are shown in Figure 5b; the solid curve represents the exact analytical expression of $D_q$ for the synthetic multifractal Sierpinski triangle of

Figure 4, the dashed curve is the estimation and the gray band corresponds to ±1σ obtained over 100 trial measurements, resulting in different configurations of circles locations. We thus check that, for this dataset, our method obtains excellent results, even for negative q values and such a small dataset.

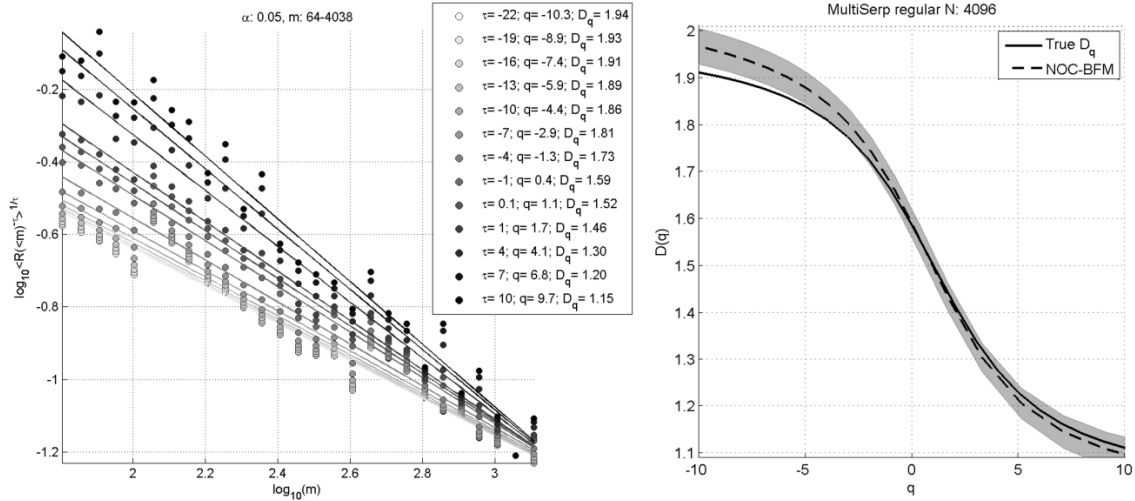

**Figure 5.** a) Averaged radii versus fixed-mass for increasing $\tau$ ; b) Analytical and estimated $D_q$-q curves for the multifractal Sierpinski triangle obtained recursively with the density generator [1 0; 1 2] as explained in the text and with Error! Reference source not found..

### *3.2 Multifractal spectra of synthetics datasets*

This subsection extends the previous one by exploring the merits and limitations of our barycentric fixed mass method applied to different mono- and multifractal measures obtained synthetically. The synthetic datasets are constructed by recursive replication of a 2 by 2 density matrix (See Figure 3 for an example). The resulting density grid is normalized so that the lowest mass in a cell is 1 and each other grid cell is uniformly sampled by a number of points according to its computed mass. An optimal uniform distribution within grid cells featuring more than one point is ensured by assessing the locations according to an optimal circle packing within each square. The analytical values of $D_q$ of the obtained distribution are given by Equation (4) in terms of the elements of the density matrix ($p_i$):

$$D_q = \frac{1}{1-q} \frac{\log(\sum_{i=1}^{4} p_i^q)}{\log(2)} \qquad (4)$$

The formulation of $D_q$ does not depend on the locations of the $p_i$ elements, thus shuffling the elements of the density matrix would result in a different fractal with the same $D_q$. To test the robustness of the new method, we conduct two test cases for each density matrix: (i) a regular distribution where the density matrix is constant through all iterations, and (ii) a random distribution where the density matrix is shuffled in the beginning and rotated 90 degrees after each iteration. We choose rotation rather than shuffling in order to ensure that the replication is not done with the same matrix in consecutive generations (a possibility when shuffling). This is done to minimize the effect of discrete scale invariance, which is intrinsic in deterministic synthetic fractals [22]. The 6 test-sets and their details are given in Table 1. The figures show the density of data points, which increases with the darkness of the grey level.

**Table 1.** Synthetic datasets used for the benchmark test

| Name | Density Matrix | Replication No | Total Mass | Density Grid | |
|---|---|---|---|---|---|
| | | | | Regular | Random |
| Monofractal Sierpinski Triangle | $\begin{bmatrix} \frac{1}{3} & 0 \\ \frac{1}{3} & \frac{1}{3} \end{bmatrix}$ | 8 | 6561 ($3^8$) | 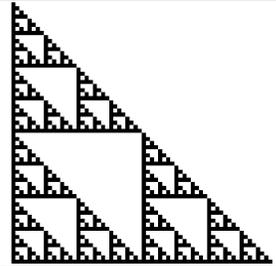 | 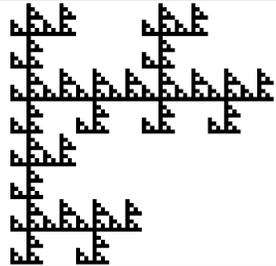 |
| Multifractal Sierpinski Triangle | $\begin{bmatrix} \frac{1}{4} & 0 \\ \frac{1}{4} & \frac{2}{4} \end{bmatrix}$ | 6 | 4096 ($4^6$) | 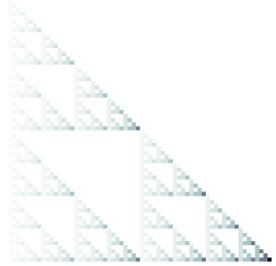 | 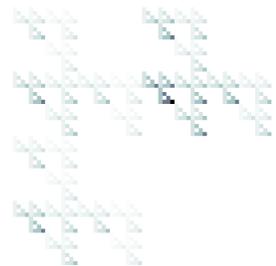 |
| Multifractal Sierpinski Carpet | $\begin{bmatrix} \frac{2}{5} & \frac{1}{5} \\ \frac{1}{5} & \frac{1}{5} \end{bmatrix}$ | 5 | 3125 ($5^5$) | 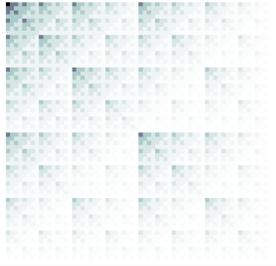 | 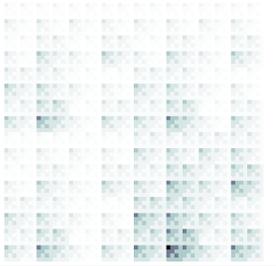 |

The proposed non-overlapping barycentric fixed-mass method (NO-BFM) was benchmarked against the usual fixed-sized (FS-SB) and fixed-mass sandbox (FM-SB) methods, both of these methods being prone to significant edge effects: the mass vs radius (or vice versa) growth for a point located at some edge of the fractal differs significantly from a point in the center of the fractal, as the corresponding circles include more and more empty space as their radius increases.

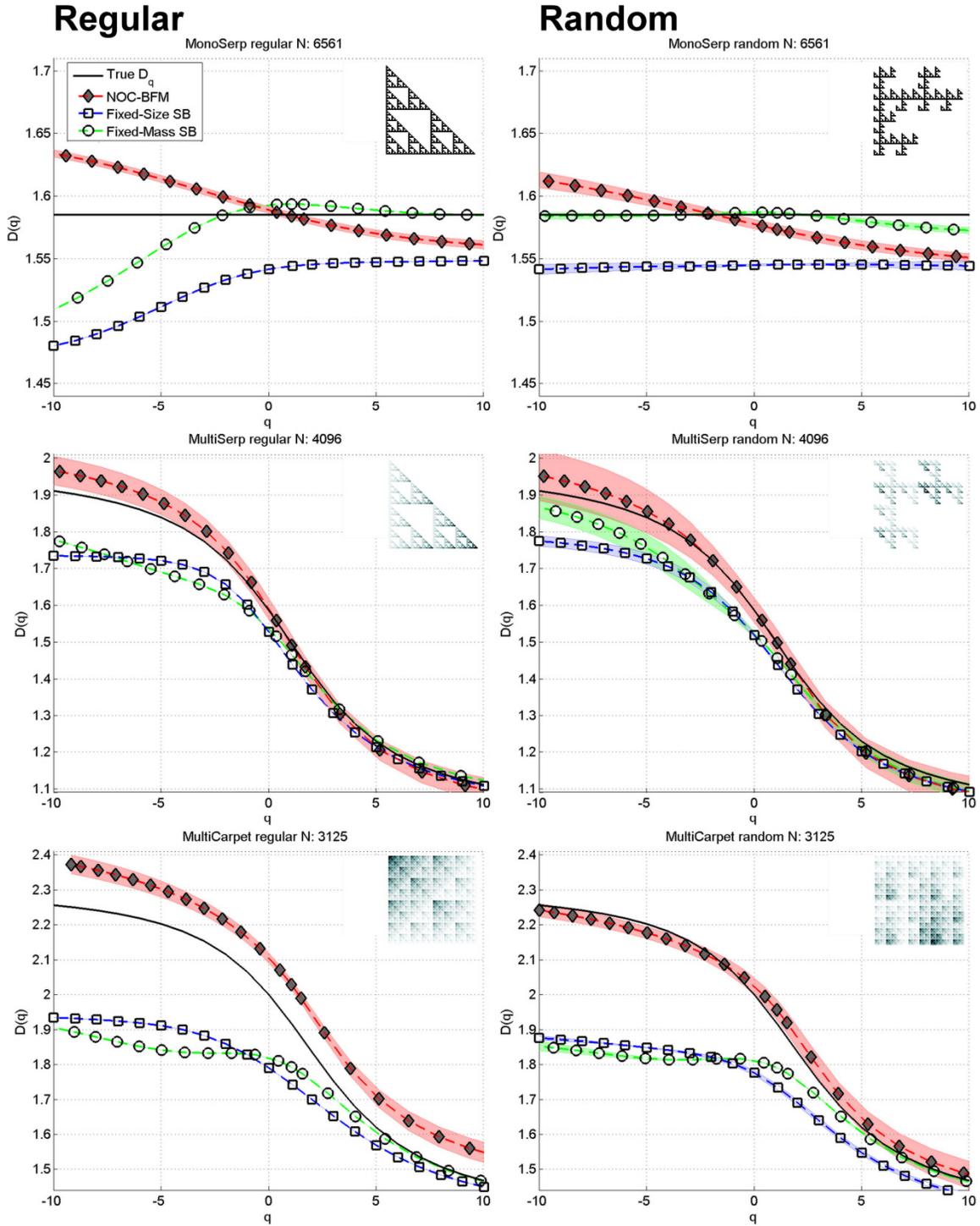

**Figure 6.** Benchmark results for regular and random synthetic distributions

In order to present an objective comparison, we tried to optimize the implementation and maximize the performance of all methods in competition. For the FS-SB method, we limited the spanning radius range starting from the maximum of the closest neighbour distance up to the minimum of the furtherest neighbour distance (corresponding to the smallest enclosing radius). This $R$ interval was logarithmically sampled as $R_i = R 10^{i\alpha}$ with

α=0.05. For the FM-SB method, the mass range $m$ was sampled logarithmically as $m_i=m10^{i\alpha}$ with α=0.05. The minimum mass was set by a condition imposed on the maximum number of points within an elementary cell, so as to avoid that the spatial distribution of points within that cell becomes uniform (D=2), thus breaking self-similarity. The maximum mass was limited to one quarter of the total mass ($1/2^D$ with D=2). These values were also used for the NO-BFM method with the exception of the maximum mass. The latter is determined automatically as the maximum mass level at which only one non-overlapping circle can be placed on the fractal, since the averaged <R> term would require at least two circles. It is important to point out that FS methods are sensitive to the sampling of the $R$ range: as a result, <M> vs R curves can become unstable at oversampled small scales where the radius increment fails to result in mass increment. In contrast, the FM methods are more robust since the radius increases to enclose the $M$th closest neighbor, thus ensuring <R> to increase with $M$. Another issue regarding the benchmarking of the methods is the use of the full (or partial) dataset for the generalized multifractal dimensions estimation. The general practice for both FM-SB and FS-SB methods is to select a random sample dataset (usually 10 percent of the whole) as pivot points, and perform the measurements at these points. The standard deviation of the measure is calculated by different randomizations of this subset. For comparability with the NO-BFM method, which considers the full dataset for determining the candidate pivot points and eliminating overlaps, we also use the full dataset as pivot points for both FM-SB and FS-SB methods. Thus, for the regular fractals, these methods do not allow us to compute a standard deviation, while NO-BFM provides a standard deviation related to the different possible packing configurations.

The $D_q$ estimates of the six datasets for the three methods are displayed in Figure 6. For the random synthetics, estimates are plotted with confidence bounds of ±1σ obtained over 100 different randomized distributions. One should be aware that all the possible randomization outcomes depend on the unique elements of the density matrices. The results clearly show that, for the multifractal sets, NO-BFM outperforms both fixed-mass and fixed-sized methods for both negative and positive values of $q$. We observe that, for the regular multifractals, the method tends to slightly overestimate $D_q$. Since we do not observe this in the randomized distributions, we conclude that this overestimation is due to the discrete scale invariance effect [22], which becomes more pronounced as the high density mass is progressively being concentrated in one part of the fractal. We have also investigated individual coverage configurations of the regular multifractals, and observe sudden drops in the number of covering circles as their mass is increased, a signature of discrete scale invariance.

In terms of computation resources, although NO-BFM includes both pivot selection (BFM) and no-overlap (NOC) criteria, it is still superior to both methods. The computation of the averaging term is significantly accelerated since NOC decreases the number of averaged points as the fixed-mass increases. On the other hand, BFM limits the number of points considered in NOC, minimizing the time needed for its computation.

## 4-Application to the Diffusion-Limited Aggregation (DLA) process

Having evaluated the accuracy and precision of the proposed NO-BFM method, we now revisit the diffusion-limited aggregation (DLA) growth process, which has been the subject of many fractal analyses. Diffusion-limited aggregation occurs when particles following a random walk stick to a stationary seeding point, becoming themselves locations for the attachment of other later incoming particles, leading to the growth of a complex aggregate. The growth of such an aggregate is governed by branch (finger) formations and the consequent screening effects of channels between the fingers. Due to its simple mechanism and widespread occurrence in natural phenomena, the DLA process has been studied extensively. However analyses of its fractal properties resulted in different conclusions: some studies [23–26] suggest that the DLA cluster is a monofractal set with a constant $D_q$, independent of q while others [27–31] propose that it is a multifractal. The differing findings are likely to be influenced by the finite size and boundary effects, which affect all $D_q$ estimation methods to different degrees. In a recently published study [32], the authors tried to address this and several other issues by limiting the pivot point selection to points within a distance $0.5R_g < d < 1.5R_g$ from the cluster seed point, where $R_g$ is the radius of gyration defined as:

$$R_g^2 = \frac{1}{N}\sum_{i=1}^{N}(r_i - r_{mean})^2 \qquad (5)$$

where $r_i$ is the distance of the $i$-th particle to the barycenter of the DLA cluster.

The R interval for the slope estimation was chosen as $0.032R_g < R < 0.32R_g$. The study [32] suggests that the multifractality of the DLA cluster is less pronounced than previously thought by some authors, proposing a constant $D_q$ value of 1.69. It is difficult to assess the consistency of these results, as they are likely to be affected by the somewhat arbitrary criterion for pivot point selection and R interval. The authors also indicate that their estimate of $D_q$ is likely to be underestimated due to the boundary effects inherent in the fixed-size sandbox method.

We performed a multifractal analysis on DLA clusters using our new NO-BFM method. For this purpose, we grew 100 off-lattice DLA clusters up to a total mass of $10^6$ points each. To study the evolution of $D_q$ as the clusters grow, we estimated $D_q$ at mass increments of $10^5$ points, resulting in 10 estimates for each cluster as it grows. For robustness of the analysis, as with the synthetics, we required a minimum of 2 circles to calculate the averaging term. The mass was sampled with a logarithmic step $\alpha=0.05$. The sampling is initiated at a mass of 7 points based on the fact that one circle surrounded by six other circles corresponds to the hexagonal lattice, which is the lattice arrangement of circles with the highest density, as proven by Lagrange. The $D_q$-q curves for the 10 different masses are shown in Figure 7, where the error bars indicate the standard deviation obtained over the 100 clusters. To check the convergence of our results in Figure 8, we plot $D_q$ as a function of the number of data points in the cluster for $q=-10, 0$ and 10.

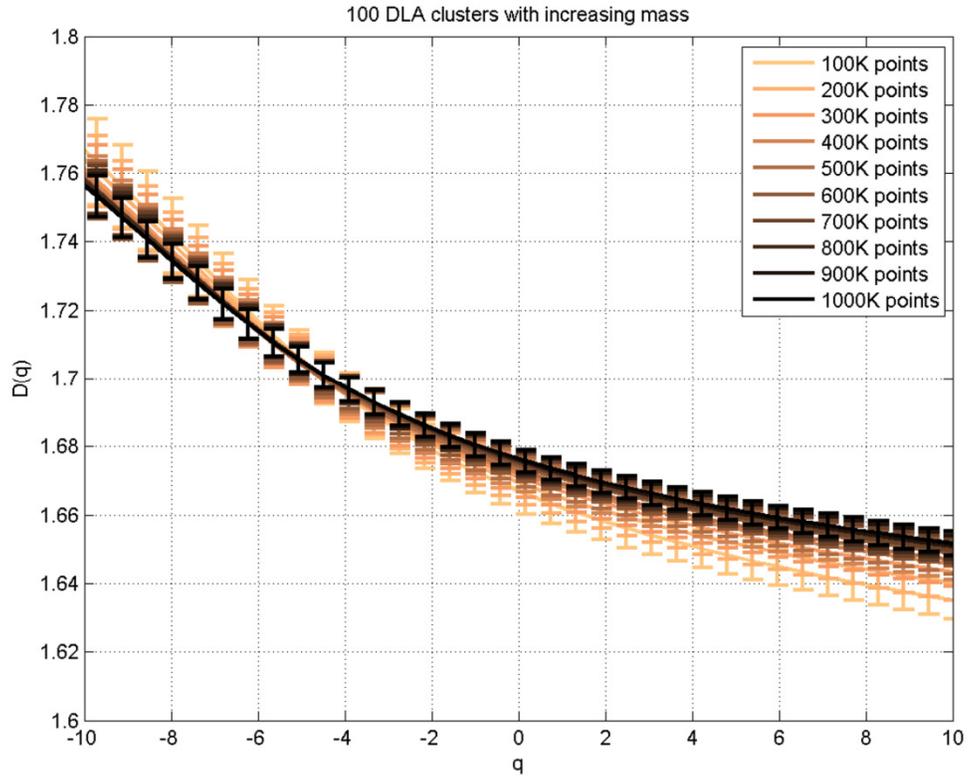

**Figure 7.** $D_q$-$q$ curves estimated for 100 DLA clusters with increasing masses given in the inset with different colors

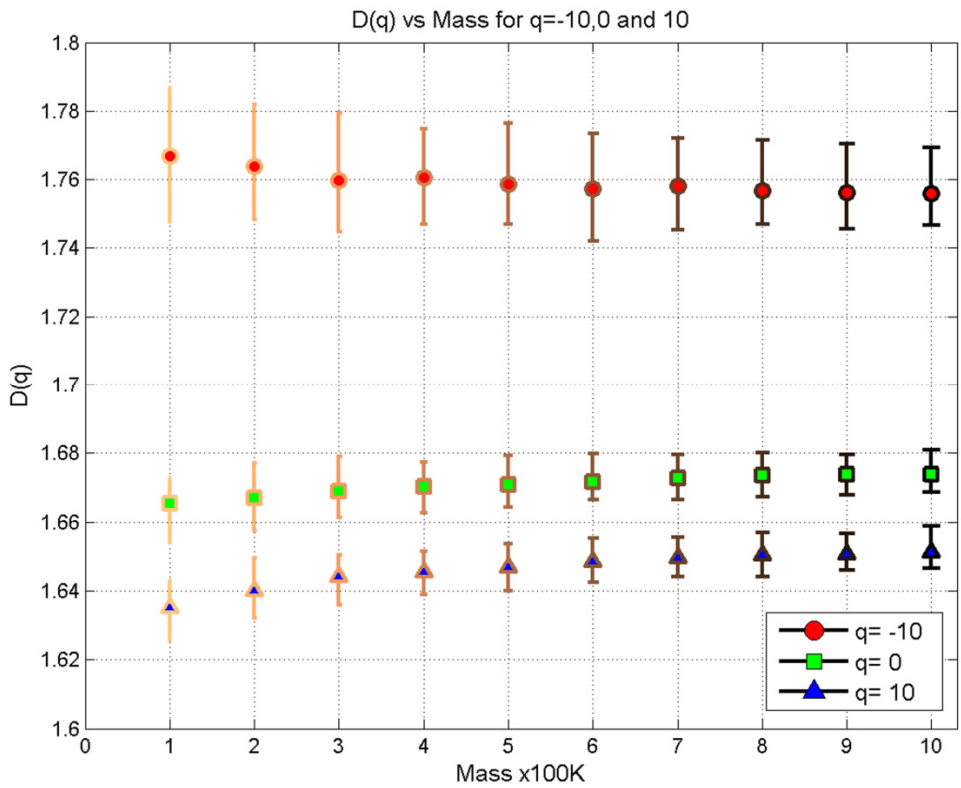

**Figure 8.** Convergence of $D_q$ versus the number of samples in DLAs

To assess the significance of the obtained measure, we conducted the following test: we created a synthetic monofractal set with $D_q=1.63$ (log(6)/log(3)) and confirmed that the method retrieves the correct dimension with a maximum error of 0.03 with sample sizes as low as $6^6=46656$ points. Secondly, we created a synthetic multifractal having a $D_q$ curve similar to what we observe in the DLA datasets. The density matrix was chosen as [2 0 2; 0 2 0; 2 1 2]. We confirmed that, for a sample size of $11^5=161051$ points, the method retrieves the correct dimensions with a maximum error of 0.02. The results are presented in Figure 9. We chose regular constructions because they can be associated with the mean DLA branching number 5, resulting in synthetic clusters that look somewhat similar to DLA clusters.

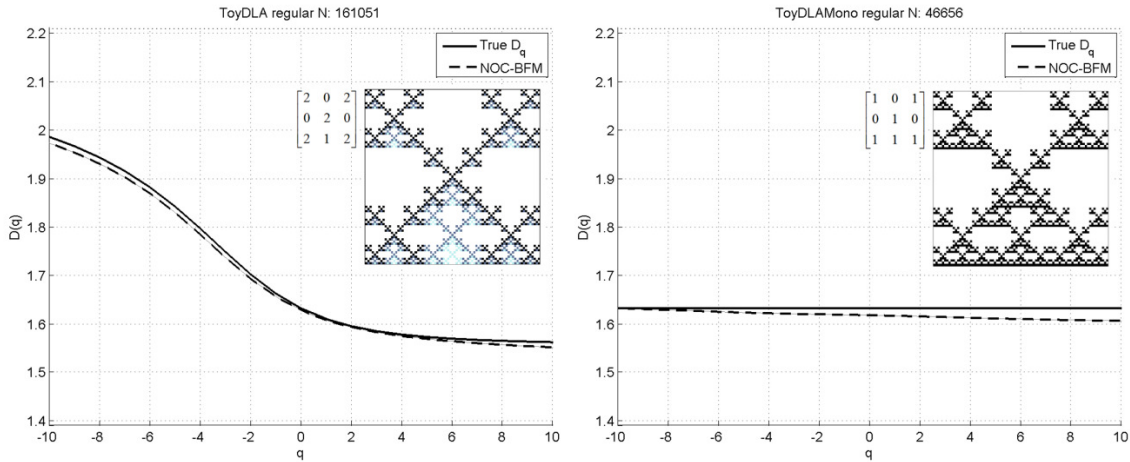

**Figure 9.** $D_q$-$q$ curves obtained for mono and multifractal toy DLA synthetics

In order to put our results in perspective, Figure 10 shows our obtained multifractal $D_q$ curves together with those obtained in [32] and [33] for DLA clusters with similar mass. All three results are for DLA clusters of 1 million particles. In contrast to [32] that uses only $10^4$ random pivot points, [33] uses $5*10^4$ random pivot points and square boxes for the fixed-size sandbox method. Also, the R interval for this study was wider: $0.037 R_{max} < R < 0.5 R_{max}$.

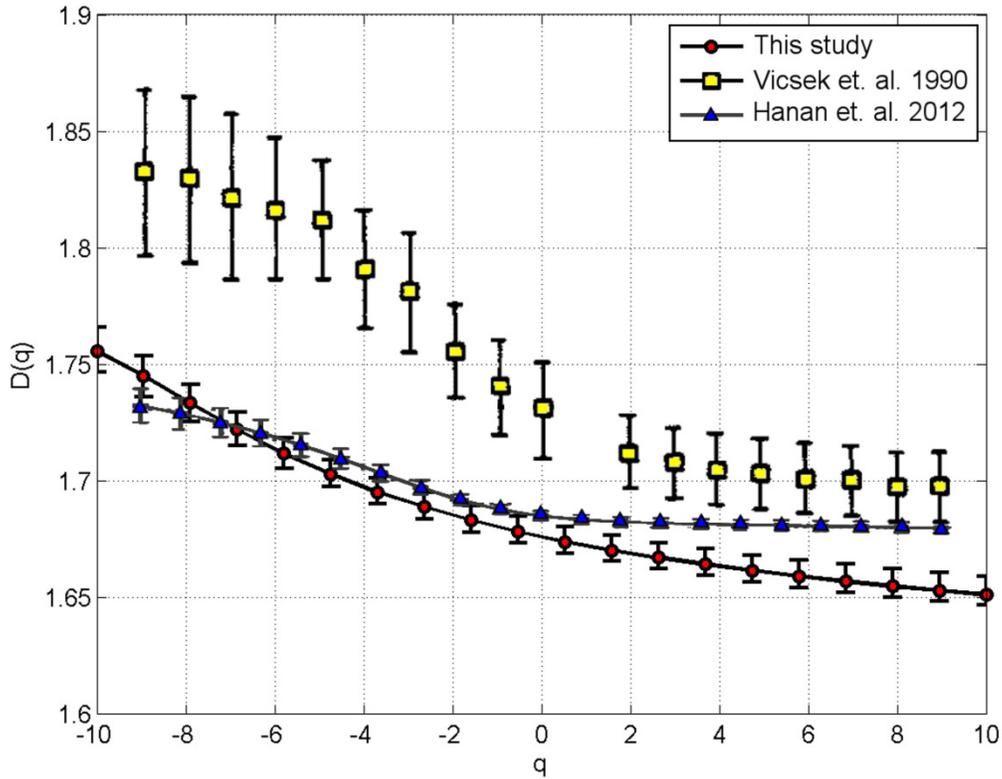

**Figure 10.** Comparison with $D_q$-$q$ curves reported by Hanan et. al. 2012 Vicsek et. al. 1990

## 5-Conclusion

We have introduced a novel method, coined the Barycentric Fixed Mass Method, to address the ubiquitous edge and finite size effects that plague current determination of multifractal spectra. The method incorporates two motivated criteria (the barycentric pivot point selection and the non-overlapping coverage) in order to reduce edge effects, improve precision and reduce computation time. We have presented extensive tests on synthetically generated mono- and multifractals, both deterministic and random, which demonstrate the superior performance of our proposed method. We have then applied it to the still open question of whether large clusters generated by diffusion-limited aggregation (DLA) exhibit genuine multifractality. Our tests support the conclusion of a genuine by weak multifractality of the central core of DLA clusters, with $D_q$ decreasing from 1.75±0.01 for q=-10 to 1.65±0.01 for q=+10.

# References

[1] B. B. Mandelbrot, *Fractals: Form, Chance and Dimension* (W.H.Freeman & Company, 1977).

[2] A. Renyi, *Probability Theory* (North-Holland, Amsterdam, 1970).

[3] G. Ouillon and D. Sornette, Geophysical Research Letters **23**, 3409 (1996).

[4] D. Weitz and M. Oliveria, Physical Review Letters **52**, 1433 (1984).

[5] M. Tence, J. P. Chevalier, and R. Jullien, Journal De Physique **47**, 1989 (1986).

[6] A. Torcini, A. Politi, P. Puccioni, and G. D'Alessandro, Physica D **53**, 85 (1991).

[7] G. Ouillon, D. Sornette, and C. Castaing, Nonlinear Processes in Geophysics **2**, 158 (1995).

[8] P. Grassberger and I. Procaccia, Physica D: Nonlinear Phenomena **9**, 189 (1983).

[9] V. H. Márquez-Rámirez, F. A. Nava Pichardo, and G. Reyes-Dávila, Pure and Applied Geophysics **169**, 2091 (2012).

[10] M. Jensen, L. Kadanoff, and A. Libchaber, Physical Review Letters **55**, 2798 (1985).

[11] T. Tél, Á. Fülöp, and T. Vicsek, Physica A **159**, 155 (1989).

[12] H. Greenside, A. Wolf, J. Swift, and T. Pignataro, Physical Review A **25**, 3453 (1982).

[13] P. Grassberger, R. Badii, and A. Politi, Journal of Statistical Physics **51**, 135 (1988).

[14] R. Badii and G. Broggi, Physics Letters A **131**, 339 (1988).

[15] T. Hirabayashi, K. Ito, and T. Yoshii, Pure and Applied Geophysics **138**, 591 (1992).

[16] J. Theiler, Journal of the Optical Society of America A **7**, 1055 (1990).

[17] A. Arneodo, G. Grasseau, and M. Holschneider, Physical Review Letters **61**, 2281 (1988).

[18] G. Ouillon, D. Sornette, and C. Castaing, Nonlinear Processes in Geophysics **2**, 158 (1995).

[19] C. Tricot, Mathematical Proceedings of the Cambridge Philosophical Society **91**, 57 (1982).

[20] F. Hausdorff, Mathematische Annalen **68**, (1918).

[21] S. Lynch, *Dynamical Systems with Applications Using MATLAB* (Birkhauser, 2004).

[22] D. Sornette, Physics Reports **297**, 239 (1998).

[23] P. Meakin, Physical Review Letters **51**, 1119 (1983).

[24] F. Argoul, A. Arneodo, G. Grasseau, and H. Swinney, Physical Review Letters **61**, 2558 (1988).

[25] G. Li, L. Sander, and P. Meakin, Physical Review Letters **63**, 2558 (1989).

[26] F. Argoul, A. Arneodo, and J. Elezgaray, Physical Review A **41**, (1990).

[27] J. Nittmann, H. Stanley, E. Touboul, and G. Daccord, Physical Review Letters **58**, 1987 (1987).

[28] P. Meakin and S. Havlin, Physical Review A **36**, (1987).

[29] T. Nagatani, Physical Review A **38**, 2632 (1988).

[30] T. Vicsek, Physica A **168**, 490 (1990).

[31] H. Boularot and G. Albinet, Physical Review E **53**, 5106 (1996).

[32] W. G. Hanan and D. M. Heffernan, Physical Review E **85**, 021407 (2012).

[33] T. Vicsek, F. Family, and P. Meakin, Europhysics Letters **12**, 217 (1990).